\documentclass[aps,prl,twocolumn,showpacs,groupedaddress,floatfix,
               amsmath,amssymb]{revtex4-1}

\usepackage{graphicx}%
\usepackage{dcolumn}
\usepackage{natbib}
\usepackage{bm}

\def\urltilda{\kern -.15em\lower .7ex\hbox{\~{}}\kern .04em}

\begin{document}

\title{Computation of Casimir Interactions between 
       Arbitrary 3D Objects with Arbitrary Material Properties}

%!!!!!!!!!!!!!!!!!!!!!!!!!!!!!!!!!!!!!!!!!!!!!!!!!!
%! authors section
%!!!!!!!!!!!!!!!!!!!!!!!!!!!!!!!!!!!!!!!!!!!!!!!!!!
\author{M. T. Homer Reid$^{1,2}$%
        \footnote{Electronic address: \texttt{homereid@mit.edu}}%
        \footnote{URL: \texttt{http://www.mit.edu/\urltilda homereid}},
       Jacob White$^{2,3}$,
       and Steven G. Johnson$^{2,4}$}
\affiliation{ $^{1}$Department Of Physics, 
              Massachusetts Institute of Technology, 
              Cambridge, MA 02139, USA \\
              $^{2}$Research Laboratory of Electronics,
              Massachusetts Institute of Technology, 
              Cambridge, MA 02139, USA \\
              $^{3}$Department of Electrical Engineering and Computer Science,
              Massachusetts Institute of Technology, 
              Cambridge, MA 02139, USA \\
              $^{4}$Department Of Mathematics,
              Massachusetts Institute of Technology, 
              Cambridge, MA 02139, USA}

\date{\today}

%%%%%%%%%%%%%%%%%%%%%%%%%%%%%%%%%%%%%%%%%%%%%%%%%%%
%%%%%%%%%%%%%%%%%%%%%%%%%%%%%%%%%%%%%%%%%%%%%%%%%%%
%% abstract
%%%%%%%%%%%%%%%%%%%%%%%%%%%%%%%%%%%%%%%%%%%%%%%%%%%
%%%%%%%%%%%%%%%%%%%%%%%%%%%%%%%%%%%%%%%%%%%%%%%%%%%
\begin{abstract}
We extend a recently introduced method for computing
Casimir forces between arbitrarily--shaped metallic objects
[M. T. H. Reid \textit{et al.}, Phys. Rev. Lett. \textbf{103} 
040401 (2009)] to allow treatment of objects with arbitrary
material properties, including imperfect conductors, dielectrics, 
and magnetic materials. Our original method considered electric 
currents on the surfaces of the interacting objects; the extended 
method considers both electric \textit{and magnetic} surface 
current distributions, and obtains the Casimir energy of a configuration of 
objects in terms of the interactions of these effective surface 
currents.
Using this new technique, we present the first predictions of 
Casimir interactions in
several experimentally relevant geometries that would be 
difficult to treat with any existing method. In particular, 
we investigate Casimir interactions between dielectric 
nanodisks embedded in a dielectric fluid; we identify the 
threshold surface--surface separation at which finite--size
effects become relevant, and we map the rotational energy 
landscape of bound nanoparticle diclusters.
\end{abstract}

\maketitle

%%%%%%%%%%%%%%%%%%%%%%%%%%%%%%%%%%%%%%%%%%%%%%%%%%%
%%%%%%%%%%%%%%%%%%%%%%%%%%%%%%%%%%%%%%%%%%%%%%%%%%%
%% begin main text %%%%%%%%%%%%%%%%%%%%%%%%%%%%%%%%
%%%%%%%%%%%%%%%%%%%%%%%%%%%%%%%%%%%%%%%%%%%%%%%%%%%
%%%%%%%%%%%%%%%%%%%%%%%%%%%%%%%%%%%%%%%%%%%%%%%%%%%
Rapid progress in experimental Casimir 
physics~\cite{munday09} has created a demand for 
theoretical tools that can predict Casimir forces and
torques on bodies with realistic material properties,
configured in experimentally relevant geometries.
Techniques for computing Casimir interactions have traditionally
fallen into two categories~\cite{*[{Here we are considering
methods for computing Casimir interactions among bodies
of \textit{arbitrary} material properties; in the special
case of dilute gases (corresponding to relative 
permittivity close to unity), the interactions are
approximately pairwise and other computational methods
may be used; see, e.g., }] [{ }] Milton08}. In the 
\textit{scattering} approach~\cite{rahi10}, the Casimir 
energy is obtained from
spectral properties of the classical electromagnetic (EM)
scattering matrix of the geometry; this method 
leads to rapidly convergent and even \textit{analytically}
tractable~\cite{EGJK}\, formulae for certain highly symmetric 
geometries in which a special choice of coordinates
leads to a simple form for the scattering matrix, but is less
readily applicable to objects of general asymmetric shapes.  
In the alternative \textit{numerical stress--tensor} (ST) 
approach~\cite{VirtualPhotons}, the Casimir force and torque 
on a compact body are obtained by numerically integrating the 
(thermally and quantum--mechanically averaged) Maxwell ST over 
a bounding surface surrounding that body, with the value of 
the ST at each spatial point obtained by solving a classical 
EM scattering problem. The ST method has the virtue of 
flexibility, as almost any of the myriad existing 
methods for computational EM packages may be applied 
off-the-shelf to solve scattering problems for arbitrary geometries 
and materials, but the ST integration adds a layer of conceptual
and computational complexity.

In recent work~\cite{Reid09}, we have introduced a new method
for Casimir computations that combines the best features of
the scattering and stress--tensor approaches. Our 
\textit{surface--current} technique discretizes the surfaces
of interacting objects into small surface patches 
(Fig. 1)---thus retaining the full flexibility of 
the numerical ST approach in handling arbitrarily
complex, asymmetric geometries---but bypasses the unwieldy
ST integration by obtaining the Casimir energy 
directly from spectral properties of a matrix describing the 
interactions of \textit{surface currents} on the
discretized object surfaces---thus mirroring the conceptual 
directness of the scattering approach, 
%%%%%%%%%%%%%%%%%%%%%%%%%%%%%%%%%%%%%%%%%%%%%%%%%%
%%%%%%%%%%%%%%%%%%%%%%%%%%%%%%%%%%%%%%%%%%%%%%%%%%
% figure 1
%%%%%%%%%%%%%%%%%%%%%%%%%%%%%%%%%%%%%%%%%%%%%%%%%%
%%%%%%%%%%%%%%%%%%%%%%%%%%%%%%%%%%%%%%%%%%%%%%%%%%
\begin{figure}[h]
\resizebox{8.6cm}{!}{\includegraphics{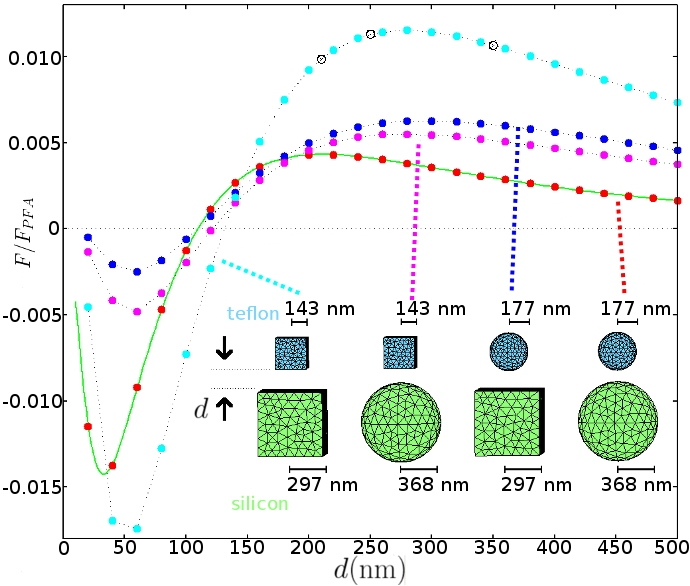}}
\caption{Casimir force between silicon and teflon 
spheres and cubes immersed in ethanol, normalized 
by the PFA predictions (see text). For
the sphere-sphere case, the solid circles represent
data computed using the methods described in this
paper, while the thick solid line reproduces data
from Ref.~\cite{diclusters}. For the other three cases,
the solid and hollow circles represent data computed 
using the methods described in this paper, and the dotted 
lines are a guide to the eye.}
\end{figure}
%%%%%%%%%%%%%%%%%%%%%%%%%%%%%%%%%%%%%%%%%%%%%%%%%%
dramatically improving computational efficiency 
for moderate-size problems where dense linear algebra
is applicable, and establishing
a connection to the equivalent-surface-current method
of classical EM~\cite{Harrington2001}.

The technique discussed in~\cite{Reid09} was
limited to the case of perfect-metal bodies in vacuum. In this 
paper we extend the method to allow 
treatment of materials with \textit{arbitrary} frequency-dependent
(but spatially constant) permittivity $\epsilon$ and permeability 
$\mu$. A thorough description of our extended formulation
will be provided elsewhere~\cite{Reid10B}; here we present a 
brief overview of our method and demonstrate its use in predicting 
Casimir interactions in systems that would be difficult to treat by 
any other method.

\textit{Casimir Interactions from Surface Currents.}
The Casimir energy of a collection of objects is given quite generally
by an expression of the form \footnote{We work here at 
zero temperature $T$; the extension to finite $T$ is straightforward
and will be presented in~\cite{Reid10B}.}
\begin{equation}
   \mathcal{E}=-\frac{\hbar}{2\pi} 
   \int_0^\infty \, d\xi \, \log \frac{\mathcal{Z}(\xi)}{\mathcal{Z}_\infty(\xi)} 
 \label{Eexpr} 
\end{equation}
where the partition function for the electromagnetic field
at imaginary frequency $\xi$ is 
\begin{equation}
  \mathcal{Z}(\xi)=\int \left[\mathcal{D}\mathbf{A} \right]_c
  e^{-\frac{1}{2}\int 
      \mathbf{A} (\xi, \mathbf{x}) 
      \cdot 
      \boldsymbol{\mathcal{G}}^{-1}(\xi, \mathbf{x},\mathbf{x}^\prime) 
      \cdot 
      \mathbf{A}(\xi, \mathbf{x}^\prime)
      \, d\mathbf{x} \,  d\mathbf{x}^\prime 
    };
 \label{Zexpr}
\end{equation}
here $\boldsymbol{\mathcal{G}}$ is the (gauge-fixed)
photon propagator, and the notation $[\cdots]_c$ indicates 
that this is a \textit{constrained} path integral, in which 
the path integration ranges only over field configurations 
$\mathbf{A}$ that satisfy the boundary conditions in the 
presence of the interacting objects. 
($\mathcal{Z}_\infty$ in (\ref{Eexpr}) is $\mathcal{Z}$
with all objects removed to infinite separations.) 

Although the exponent in (\ref{Zexpr})
is only quadratic in the integration variables, the implicit 
constraint in the integration measure prevents direct evaluation
of the path integral. We obtain a more tractable expression
by representing the constraints on the $\mathbf{A}$ field 
\textit{explicitly} via functional $\delta$ 
functions~\cite{Bordag85, Kardar91}.
Consider a single point $\mathbf x$ on the surface of an
object in our geometry. The boundary conditions 
at $\mathbf x$ are that the components of $\mathbf E$ and $\mathbf H$ 
tangential to the object surface are continuous as we pass
from the interior to the exterior of the object at $\mathbf x$:
$$ \mathbf E^{\text{\scriptsize{in}}}_{\parallel}(\mathbf x) 
  =\mathbf E^{\text{\scriptsize{out}}}_{\parallel}(\mathbf x), \qquad
   \mathbf H^{\text{\scriptsize{in}}}_{\parallel}(\mathbf x)
  =\mathbf H^{\text{\scriptsize{out}}}_{\parallel}(\mathbf x).
$$
We introduce two 2-dimensional $\delta$ functions that impose 
these constraints at $\mathbf{x}$:
{\small
%====================================================================%
\begin{align*}
%--------------------------------------------------------------------%
 \delta
   \Big( \mathbf E^{\text{\scriptsize{in}}}_{\parallel}(\mathbf x) 
        -\mathbf E^{\text{\scriptsize{out}}}_{\parallel}(\mathbf x)
   \Big)
&=\int_{-\infty}^{\infty} 
  \int_{-\infty}^{\infty} \frac{dK_1 dK_2}{(2\pi)^2}
  e^{i\mathbf K\cdot[\mathbf E^{\text{\tiny{in}}}(\mathbf x) -
                     \mathbf E^{\text{\tiny{out}}}(\mathbf x)]}
\\
%--------------------------------------------------------------------%
 \delta
   \Big( \mathbf H^{\text{\scriptsize{in}}}_{\parallel}(\mathbf x) 
        -\mathbf H^{\text{\scriptsize{out}}}_{\parallel}(\mathbf x)
   \Big)
&=\int_{-\infty}^{\infty} 
  \int_{-\infty}^{\infty} \frac{dN_1 dN_2}{(2\pi)^2}
  e^{i\mathbf N\cdot[\mathbf H^{\text{\tiny{in}}}(\mathbf x) -
                     \mathbf H^{\text{\tiny{out}}}(\mathbf x)]}
\end{align*}
}
%====================================================================%
\noindent
where we may think of the Lagrange multipliers $\mathbf K$ and 
$\mathbf N$ as two-component vectors living in the tangent 
space to the object surface at $\mathbf x$.
Aggregating the corresponding $\delta$ functions 
for \textit{all} points $\mathbf x$ on all object surfaces, we 
obtain an explicit \textit{functional} $\delta$ function 
constraining the electromagnetic field to satisfy the boundary 
conditions:
\begin{equation}
 \delta\Big[ \mathbf A\Big]=
   \int \mathcal{D}\mathbf K \int \mathcal{D}\mathbf N
  \,
   e^{ \oint
       \big\{ 
       i\mathbf{K}\cdot \mathcal{L}^{\text{\tiny{E}}}\mathbf A   
      +i\mathbf{N}\cdot \mathcal{L}^{\text{\tiny{H}}}\mathbf A
       \big\}\,d\mathbf x
     }
 \label{functionaldelta}
\end{equation}
where the integration in the exponent is over the surfaces of 
all objects in our geometry, the path integrations range 
over all possible tangential vector fields 
$\mathbf K(\mathbf x), \mathbf N(\mathbf x)$ on the object 
surfaces, and 
$\mathcal{L}^{\text{\tiny{E}}},\mathcal{L}^{\text{\tiny{H}}}$ 
are differential operators that act on $\mathbf A$ to yield 
$\mathbf{E}$ and $\mathbf{H}$. Because $\mathbf{K}$ 
and $\mathbf{N}$ respectively enforce the boundary conditions
on the \textit{electric} and \textit{magnetic} fields, it is 
tempting to interpret these tangential vector fields as 
\textit{electric} and \textit{magnetic} surface current 
distributions on the surfaces of the interacting objects.
Effective surface current distributions of this sort are ubiquitous
in equivalence-principle techniques for EM field 
computation~\cite{Harrington2001} and form the basis of the 
boundary-element method of computational 
EM~\cite{PMCHW1}.

Inserting the explicit constraint (\ref{functionaldelta}) into
the path integral (\ref{Zexpr}) allows us to relax the implicit
constraint $[\cdots]_c$, whereupon the path integral over the 
$\mathbf A$ field may be performed analytically:
\begin{align}
  \mathcal{Z}(\xi)
&=\int \mathcal{D}\mathbf{A}\, 
       \mathcal{D}\mathbf{K}\, 
       \mathcal{D}\mathbf{N}\,
  e^{-\frac{1}{2}\int \mathbf{A} \boldsymbol{\mathcal{G}}^{-1}\mathbf{A}
      +i\oint \mathbf{A} \cdot[ \mathcal{L}^{\text{\tiny{E}}} \mathbf{K}
                              +\mathcal{L}^{\text{\tiny{H}}} \mathbf{N}
                             ]
    }\nonumber \\
&\sim
  \int 
  \mathcal{D}\mathbf{K}\, \mathcal{D}\mathbf{N}\,
  \exp\{-S_\xi [\mathbf{K}, \mathbf{N}]\}
 \label{Zexpr2}
\end{align}
where $\sim$ denotes equality up to constant factors that cancel
in the ratio in (\ref{Eexpr}), and where the effective action
%====================================================================%
$$ S_\xi[\mathbf{K}, \mathbf{N}]
   =
  \oint \left\{
   \binom{\mathbf{K}}{\mathbf{N}}^{\text{\scriptsize{T}}}
   \cdot
%====================================================================%
   \left(\begin{array}{cc}
%--------------------------------------------------------------------%
   \mathcal{L}^{\text{\tiny{E}}}
   \boldsymbol{\mathcal{G}}
   \mathcal{L}^{\text{\tiny{E}}} &
%--------------------------------------------------------------------%
   \mathcal{L}^{\text{\tiny{E}}}
   \boldsymbol{\mathcal{G}}
   \mathcal{L}^{\text{\tiny{H}}} \\
%--------------------------------------------------------------------%
   \mathcal{L}^{\text{\tiny{H}}}
   \boldsymbol{\mathcal{G}}
   \mathcal{L}^{\text{\tiny{E}}} &
%--------------------------------------------------------------------%
   \mathcal{L}^{\text{\tiny{H}}}
   \boldsymbol{\mathcal{G}}
   \mathcal{L}^{\text{\tiny{H}}} \\
%--------------------------------------------------------------------%
   \end{array}\right)
  \cdot
   \binom{\mathbf{K}}{\mathbf{N}} 
  \right\} d\mathbf x
%====================================================================%
$$
describes the interactions of electric and magnetic surface currents 
on the bodies of the interacting objects. 
The frequency-dependent material parameters $\epsilon(\xi), \mu(\xi)$ 
of all interacting objects, and of the medium in which they are embedded, 
enter into $S_\xi$ through 
$\mathcal{L}^{\text{\tiny{E}}}, \mathcal{L}^{\text{\tiny{H}}}$, and
$\boldsymbol{\mathcal{G}}$.

\textit{Discretization.} To evaluate (\ref{Zexpr2}) for 
a given collection of objects, we now proceed as in~\cite{Reid09}
by discretizing the surfaces of the objects into small planar
triangles~\cite{GMSH}. We
introduce a set~\cite{RWG} of localized vector-valued basis 
functions $\{\mathbf f_\alpha\}$ 
and approximate the surface currents 
$\mathbf K(\mathbf x), \mathbf N(\mathbf x)$ as expansions
in the finite set $\{\mathbf f_\alpha\}$:
$$
\mathbf{K}(\mathbf{x})
  \approx \sum_\alpha K_\alpha \mathbf{f}_\alpha(\mathbf{x}), 
  \qquad 
   \mathbf{N}(\mathbf{x})
  \approx \sum_\alpha N_\alpha \mathbf{f}_\alpha(\mathbf{x}).
$$
The functional integral in (\ref{Zexpr2}) becomes a
finite-dimensional Gaussian integral, 
\begin{align}
\int \mathcal{D}\mathbf{K}\, \mathcal{D}\mathbf{N}\,
 e^{-S_\xi [\mathbf{K}, \mathbf{N}]}
&\sim
  \int \prod dK_\alpha dN_\alpha 
   e^{-\tbinom{K}{N}
         \cdot 
         \mathbf M(\xi)  
         \cdot 
         \tbinom{K}{N}
       }
\nonumber \\
&\sim [\det \mathbf M(\xi)],^{-1/2}
 \label{Zexpr3}
\end{align}
and, inserting into (\ref{Eexpr}), the Casimir energy becomes
\begin{equation}
 \mathcal{E} = \frac{\hbar}{2\pi} \int_0^\infty d\xi 
 \log\frac{\det \mathbf{M}(\xi)}{\det \mathbf{M}_\infty(\xi)}
 \label{Eexpr2}
\end{equation}
while differentiating with respect to the position of an object
gives the Casimir force on that object:
\begin{equation}
 \mathcal{F}_i = -\frac{\hbar}{2\pi} \int_0^\infty d\xi 
 \, \text{Tr} 
  \left\{ \mathbf{M}^{-1} 
  \cdot \frac{\partial\mathbf{M}}{\partial \mathbf x_i}
  \right\}.
 \label{Fexpr2}
\end{equation}
The matrix $\mathbf{M}$ describes interactions among the effective
electric and magnetic surface currents. More specifically, each 
pair of localized basis functions 
$(\mathbf f_\alpha,\mathbf f_\beta)$ corresponds to a 
2$\times$2 block of $\mathbf{M}$ of the form
%====================================================================%
$$ \mathbf{M}_{\alpha\beta} = 
%--------------------------------------------------------------------%
   \left(\begin{array}{cc} 
   \left<\mathbf f_\alpha|\mathbf G^{\text{\tiny{EE}}}|{\mathbf f_\beta}\right>
   &
%--------------------------------------------------------------------%
   \left<\mathbf f_\alpha|\mathbf G^{\text{\tiny{EM}}}|{\mathbf f_\beta}\right>
   \\[3pt]
%--------------------------------------------------------------------%
   \left<\mathbf f_\alpha|\mathbf G^{\text{\tiny{ME}}}|{\mathbf f_\beta}\right>
   &
%--------------------------------------------------------------------%
   \left<\mathbf f_\alpha|\mathbf G^{\text{\tiny{MM}}}|{\mathbf f_\beta}\right>
   \end{array}\right)
$$
where e.g. $\mathbf{G^\text{\tiny{ME}}}$ is a dyadic Green's function
representing the magnetic field due to a point electric dipole
source.

\textit{Comparison to the BEM}.
The matrix $\mathbf M$ has precisely the same form as the matrix 
that arises in the boundary-element method (BEM) for electromagnetic 
scattering problems~\cite{PMCHW1}. Indeed, once we have
implemented the computational machinery needed to evaluate 
(\ref{Fexpr2}), we could alternatively use this same machinery 
to compute Casimir forces in a different way---namely, by
numerically integrating the ST over a bounding surface, with the 
value of the ST at each spatial quadrature point
obtained by solving a BEM scattering problem~\cite{Xiong2010}.
This procedure would entail solving linear systems 
$\mathbf M \cdot \mathbf X = \mathbf B$ for multiple vectors 
$\mathbf B$, which
in practice would be handled by first decomposing $\mathbf M$ 
into triangular factors. Such a factorization step is also required
for numerical evaluation of the determinant or trace expressions 
in (\ref{Eexpr2}-\ref{Fexpr2}). However, in our formulation,
this factorization step is \textit{all} that is required, with
the Casimir energy and force then obtained immediately from the 
simple relations (\ref{Eexpr2}-\ref{Fexpr2}); in contrast, the ST
method requires extensive pre- and post-processing, in addition to 
the factorization of the BEM matrix, to evaluate the ST integral.
Thus, in addition to the conceptual simplicity of our equations 
(\ref{Eexpr2}--\ref{Fexpr2}) as compared to the ST
formulation, our method is significantly more computationally 
efficient than the ST approach, at least as long as we consider 
problems of moderate size ($N\equiv$ dim $\mathbf M\lesssim$ 10$^4$),
for which direct linear-algebra methods are feasible.

As an explicit demonstration of this computational advantage, 
the three points represented by hollow circles on the cube--cube data 
curve in Figure 1 were computed by the ST-integration method using our 
BEM code. The resulting force predictions agree with the 
predictions of (\ref{Fexpr2}), but consume some 12--24 times more
CPU time.

Moreover, although here we derived our equation (\ref{Fexpr2}) in the
context of the scattering approach to Casimir physics, the same
equation can alternatively be shown to follow directly from an
application of the BEM to stress-tensor Casimir computations, with
the spatial integral evaluated \textit{analytically}. 
Full details of these two complementary derivations will be
presented in \cite{Reid10B}.
%%%%%%%%%%%%%%%%%%%%%%%%%%%%%%%%%%%%%%%%%%%%%%%%%%
%%%%%%%%%%%%%%%%%%%%%%%%%%%%%%%%%%%%%%%%%%%%%%%%%%
% figure 2
%%%%%%%%%%%%%%%%%%%%%%%%%%%%%%%%%%%%%%%%%%%%%%%%%%
%%%%%%%%%%%%%%%%%%%%%%%%%%%%%%%%%%%%%%%%%%%%%%%%%%
\begin{figure}[b]
\resizebox{8.6cm}{!}{\includegraphics{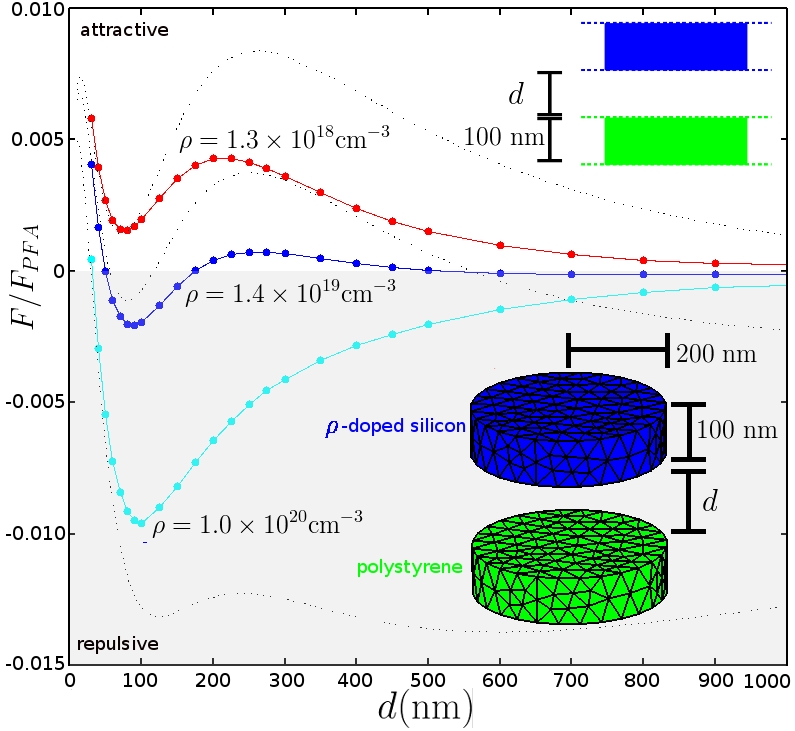}}
\caption{Casimir force between puck-shaped polystyrene
and doped-silicon (dopant density $\rho$) nanoparticles 
immersed in ethanol. The dotted curves indicate the force 
on \textit{slabs} of the same materials and thickness but 
infinite cross-sectional area~\cite{diclusters}.
(Material properties are taken from Ref.~\cite{diclusters}.)
}
\end{figure}

%%%%%%%%%%%%%%%%%%%%%%%%%%%%%%%%%%%%%%%%%%%%%%%%%%
%%%%%%%%%%%%%%%%%%%%%%%%%%%%%%%%%%%%%%%%%%%%%%%%%%
% figure 3
%%%%%%%%%%%%%%%%%%%%%%%%%%%%%%%%%%%%%%%%%%%%%%%%%%
%%%%%%%%%%%%%%%%%%%%%%%%%%%%%%%%%%%%%%%%%%%%%%%%%%
\begin{figure*}
\resizebox{17.5cm}{!}{\includegraphics{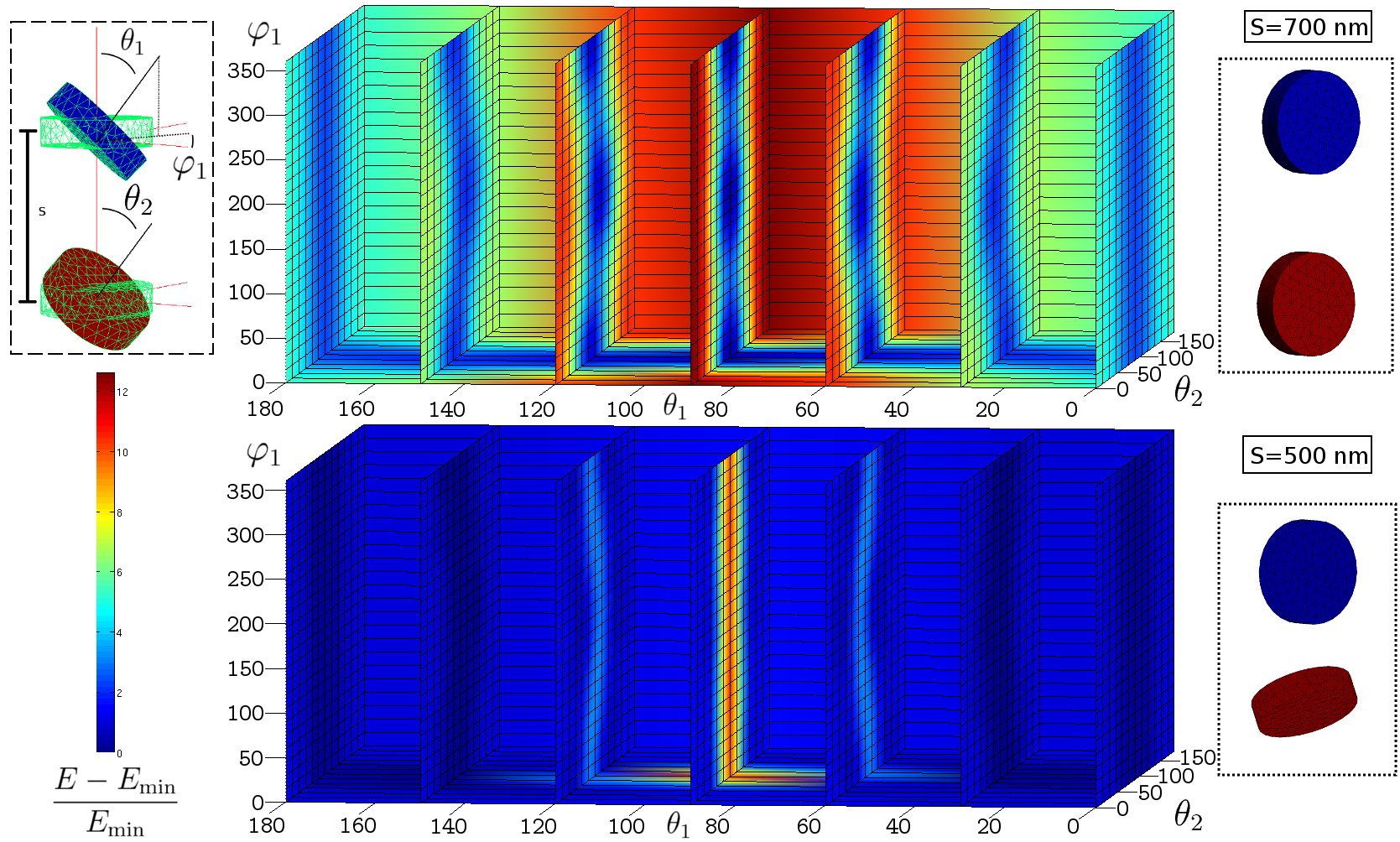}}
\caption{(Color) Rotational stability of hockey-puck shaped nanoparticles
suspended in ethanol. One nanoparticle is composed of 
polystyrene, and the other of silicon with a dopant concentration
of 1.4$\cdot 10^{19}$ cm$^{-3}$ (corresponding to the 
middle curve in the previous plot). We fix the centers of 
the particles at center--center separations of $S=\{500, 700\}$ 
nm allow them to rotate about their centers. The minimum-energy
configurations are depicted in the insets at right. (All angles are 
measured in degrees.)}
\end{figure*}

\textit{Applications.} We now use our extended method to 
predict Casimir interactions in a number of new systems that 
would be difficult to treat with any other method.

\textit{Attractive and repulsive Casimir forces between
silicon and teflon spheres and cubes immersed in ethanola.}
Ref.~\cite{diclusters}, using the
methods of Ref.~\cite{EGJK}, predicted Casimir forces between
spheres and slabs composed of various materials and suspended in 
ethanol. Figure 1 validates our method by using it to reproduce 
the results of Ref.~\cite{diclusters} for the Casimir force between
teflon and (undoped) silicon spheres, then extends these results
by plotting Casimir forces between spheres and cubes of the same
material properties (with the cubes sized to have the same 
volume as the corresponding spheres). For each geometry, we 
plot the ratio of the computed Casimir force to the force
predicted by the perfect-metal proximity force approximation 
(PFA) for that geometry; since the PFA always predicts an 
attractive force, a positive (negative) ratio implies 
an attractive (repulsive) Casimir force.
It is interesting that
all four geometries exhibit a repulsive--attractive force
crossover at approximately the same surface--surface 
separation distance.  

\textit{Finite-size effects in nanoparticle diclusters.}
Ref.~\cite{diclusters} predicted Casimir forces between
dielectric slabs, of infinite cross-sectional-area, embedded
in ethanol. It is interesting to ask how these predictions
are modified for \textit{finite} slabs (discs).
Figure 2 plots the PFA-normalized Casimir force between pairs
of puck-shaped nanoparticles, one composed of polystyrene and
the other of doped silicon, immersed in ethanol. For comparison,
the dotted curves indicate the (PFA-normalized) force between
pairs of infinite slabs of the same materials.
For small values of the surface--surface separation $d$, the 
normalized puck-puck force tends to the slab-slab limit, but 
finite-size effects cause the curves to deviate beyond
separations of $d\sim R/2$ (with $R$ the puck radius).

\textit{Rotational stability of dielectric nanoparticles.}
For the nanoparticle pair corresponding to the intermediate
curve in Figure 2 (dopant density of 
$1.4\cdot 10^{19}\text{ cm }^{-3}$ for the silicon nanoparticle),
we now fix the centroids of the nanoparticles and plot the 
Casimir energy (Figure 3) as a function of rotation angles 
(see the inset of Figure 3) for two centroid--centroid separation
distances $S$ of 700 and 500 nm. The slight shift in the value
of $S$ changes the stable equilibrium (right).

\textit{Acknowledgements.}
We are grateful to A. Rodriguez for providing the raw data
from Ref.~\cite{diclusters}.
The authors are grateful for support from the Singapore-MIT 
Alliance Computational Engineering flagship research program.

%\bibliography{PRL0710}
%\end{document}

%merlin.mbs 2010-03-15 4.21a (PWD, AO, DPC)
\providecommand{\noopsort}[1]{}\providecommand{\singleletter}[1]{#1}%

\end{document}